Alkhimova S.

# IMPACT OF PERFUSION ROI DETECTION TO THE QUALITY OF CBV PERFUSION MAP

*Об'єктом дослідження даної роботи є якість перфузійної CBV мапи з огляду визначення області перфузії як ключового компонента в процесі обробки зображень перфузійної динамічно-сприйнятливої контрастної магнітно-резонансної томографії голови людини. CBV мапа загальновизнана як найкраща для визначення локалізації та розмірів зони ураження при інсульті і в оцінюванні ангіогенезу пухлин головного мозку. Низька якість цієї мапи може спричинити хибні результати як кількісних розрахунків, так і візуальної оцінки церебрального об'єму крові.*

*Вплив визначення області перфузії на якість CBV мап був проаналізований шляхом порівняння мап, що були отримані із зображень після застосування порогової фільтрації та з еталонних зображень від 12 пацієнтів із цереброваскулярними захворюваннями. Область перфузії головного мозку була визначена вилученням пікселів низької (ділянки повітря та екстрацеребральних тканин) та високої (ділянки спинномозкової рідини) інтенсивності. Мапи були отримані методом визначення площі під кривою та методом деконволюції.*

*Для обох методів мапи, що порівнювалися, мали сильний позитивний взаємозв'язок за даними кореляційного аналізу Пірсона: r=0,8752 та r=0,8706 для методу визначення площі під кривою та методу деконволюції відповідно, $p<2,2\cdot10^{-16}$. Незважаючи на це, для обох методів діаграми розсіювання мали точки, пов'язані з втраченими зонами кровопостачання, а лінії регресії вказували на наявність помилок масштабу і зсуву для мап, що були отримані із зображень після застосування порогової фільтрації.*

*Отримані результати вказують на те, що порогова фільтрація є неефективним способом визначення області перфузії головного мозку, використання якого може спричиняти погіршення якості CBV мапи. Визначення області перфузії має бути стандартизоване та додане до протоколів перевірки нових систем аналізу перфузійних даних.*

**Ключові слова:** *перфузійна динамічно-сприйнятлива контрастна магнітно-резонансна томографія, церебральний об'єм крові, зона уваги, порогова фільтрація.*

## 1. Introduction

Analysis of the dynamic susceptibility contrast (DSC) magnetic resonance (MR) brain perfusion data is extensively used in medical practice and it is still a subject of ongoing research efforts. In DSC perfusion MR imaging, the passage of gadolinium-based contrast agent through the tissues causes a signal drop on T2-weighted images. Processing of the achieved signal intensity time curves on a pixel-by-pixel basis provides quantification of various hemodynamic parameters and uses to generate color-coded perfusion maps.

The values of cerebral blood volume (CBV) perfusion map reflect the fraction of the pixels that contain blood vessels. It is commonly accepted to be the best perfusion map to evaluate the location and size of stroke lesions and the angiogenesis of brain tumors [1, 2].

The calculation of CBV from DSC perfusion MR studies is based on the following assumptions [3]:

1) the contrast agent is retained in the intravascular space and does not diffuse into tissues;
2) stability of the flow during the measurement;
3) T1-effects to be negligible after injecting the contrast agent.

As a result, the CBV value is higher in pixels that contain only vessel data compared to those that contain a combination of vessels and brain parenchyma data.

Different software programs for DSC perfusion data analysis use different algorithms of CBV values evaluation [4, 5]. Mostly it is evaluated as a mathematical-based semi-quantitative (summary) parameter using area under the curve method or as a physiological-based quantitative parameter using deconvolution method. The first method provides calculation of, so-called, negative enhancement integral that value is roughly proportional to the CBV and can be obtained as total area described by the baseline and the signal loss due to passage of contrast agent through the tissues. The second method requires knowledge of the contrast agent concentration in the artery supplying the region of interest to take into account the variety of physiological conditions. In this method the CBV is obtained from the ratio of the areas under the tissue and arterial signal intensity time curves, respectively.

However, not only these differences in the calculation algorithms potentially impact the quality of CBV values evaluation, and, as a result, to the diagnostic efficiency of DSC perfusion MR imaging [6]. Poor detection of brain perfusion region of interest (ROI) affects both quantitative measurements and visual assessment of analyzed data [7, 8].

*Therefore, the object of research* in this study is quality of CBV perfusion map, considering detection of perfusion ROI as a key component in processing of DSC MR images of a human head.

*The aim of research* is analysis of the impact of perfusion ROI detection on the quality of CBV perfusion maps in patients with known cerebrovascular disease using both semi-quantitative and physiological-based perfusion analysis levels.

## 2. Methods of research

Despite the availability of automatic brain tissue segmentation [9–11] in most of the modern software programs for perfusion data analysis, the user-defined thresholding method is still the state-of-the-art technique that is widely used as a way of managing pixels involved in brain perfusion ROI.

Thresholding technique bases on the dividing of all image pixels into two separate classes, i.e. background and foreground. In the case of DSC perfusion data, the preprocessing step should exclude low intensity pixels (air pixels and pixels that represent non-brain tissues) and high intensity pixels (pixels that represent regions filled with cerebrospinal fluid) as background [12]. Therefore, the thresholding should be bidirectional and should provide a binary mask $M(x,y)$ for processed image as follows:

$$M(x,y) = \begin{cases} 0, & \text{if } I(x,y) \leq \min T; \\ 1, & \text{if } \max T < I(x,y) < \min T; \\ 0, & \text{if } I(x,y) \geq \max T, \end{cases} \quad (1)$$

where $I(x,y)$ – image intensity at pixel with coordinates $(x,y)$; $\min T$ and $\max T$ – threshold values for excluding low and high intensity pixels respectively.

In order to provide a tool for managing pixels involved in brain perfusion ROI the most part of the software programs has intuitive user control for threshold values selection. Fig. 1 presents the basic level scheme of such control.

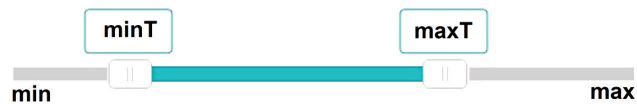

**Fig. 1.** Basic level scheme of user control for the selection of threshold values in the scope of managing pixels that are involved in brain perfusion ROI

The main limitation of such threshold-based control as a tool to manage pixels of brain perfusion ROI is that pixels having the same intensity level will always be included in the same class. For this reason, the results of thresholding technique applying are limited by overlapping the pixels intensities in different classes. In case of no overlap presence, the results can be perfect. However, this case is completely falling apart while processing clinical images. Usually, clinical images demonstrate the presence of abnormal brain anatomy with overlapping pixel intensities in lesion regions and regions which should be excluded from further perfusion analysis. From this point of view, threshold-based delineation of brain perfusion ROI can impact the quality of CBV values evaluation, which is directly affected by pixels involved in the analysis.

Quality of CBV perfusion maps, generated from threshold images, was evaluated based on 12 DSC perfusion MR datasets from 12 patients with known cerebrovascular disease. The results shown here are in whole based upon data generated by the TCGA Research Network [13].

Scan parameters for analyzed datasets were: repetition time = 1900 ms, echo time = 40 ms, flip angle = 90°, field of view = 23×23 cm, image matrix = 128×128, slice thickness = 5 mm, no gap. 5 slices were scanned with 95 dynamic images for each slice. All images from analyzed datasets were stored in 12-bit DICOM (Digital Imaging and Communication in Medicine) format.

For the purpose of analyzing the impact of perfusion ROI detection on the quality of CBV perfusion maps, the results of the applying of thresholding on the original images were used to produce semi-quantitative (area under the curve method) and physiological-based (deconvolution method) values of cerebral blood volume. Obtained values of CBV perfusion maps were pixel-wise compared with maps produced with the same methods from the reference standard.

The reference standard was defined as a manually marked ROI of the brain perfusion data by the experienced radiologist and confirmed by the second radiologist.

In order to make the comparison more meaningful, threshold values for low and high intensities exclusion were defined to present thresholding results with the highest similarity to the reference standard. For this purpose, each value from the intensities range of the analyzed image was checked via the brute force method. Each candidate of threshold values for low intensities exclusion was checked without taking into account its impact on high intensities exclusion, and vice versa.

The similarity of threshold images to the reference standard was estimated using Dice similarity index $DSI$, which value was calculated as follows:

$$DSI = \frac{2 \cdot TP}{2 \cdot TP + FP + FN}, \quad (2)$$

where $FP$ – false positive pixels that belong to the brain perfusion ROI on threshold images, but that are not associated with the same position on reference standard; $FN$ – false negative pixels that belong to the background on threshold images, but that are brain perfusion ROI pixels on reference standard; $TP$ – true positive pixels that belong to the brain perfusion ROI on threshold images and that are associated with the same position on reference standard.

## 3. Research results and discussion

The relationships between CBV perfusion maps that were produced from the threshold images and from the reference standard were investigated using Pearson product-moment correlation coefficient. Distribution of data was assessed for normality using Shapiro-Wilk test ($p > 0.05$) and a visual inspection of normal $Q$–$Q$ plot.

Two tests were conducted using different methods to produce CBV perfusion maps: one test was using area under the curve (weighted sum of the estimates of rectangle and trapezoid rule) method, and the other test was using deconvolution (block-circulant singular value decomposition) method.

The $p$-value of both tests was $< 2.2 \cdot 10^{-16}$, which is less than the significance level (alpha) of 0.05. CBV perfusion maps that were produced from the threshold images and from the reference standard are significantly correlated with Pearson correlation coefficient $r=0.8752$ (95 % CI=0.8736 to 0.8767) for the first test and $r=0.8706$ (95 % CI=0.8689 to 0.8723) for the second one (Fig. 2).

Despite the fact that results are primarily correlational, scatter plots of both tests have data points associated with missed blood regions on CBV maps that were produced from threshold images (points where y equal zero). The regression lines indicated that CBV perfusion maps produced from threshold images were subject to scale and offset errors for both tests. Overall, the relationship between analyzed CBV maps obtained for both tests was very similar. However, the slightly increased agreement was observed in case of comparison of CBV maps produced with area under the curve method.

All of the above-mentioned results agree well with a visual inspection of CBV perfusion maps produced with area under the curve method (Fig. 3) and with deconvolution method (Fig. 4).

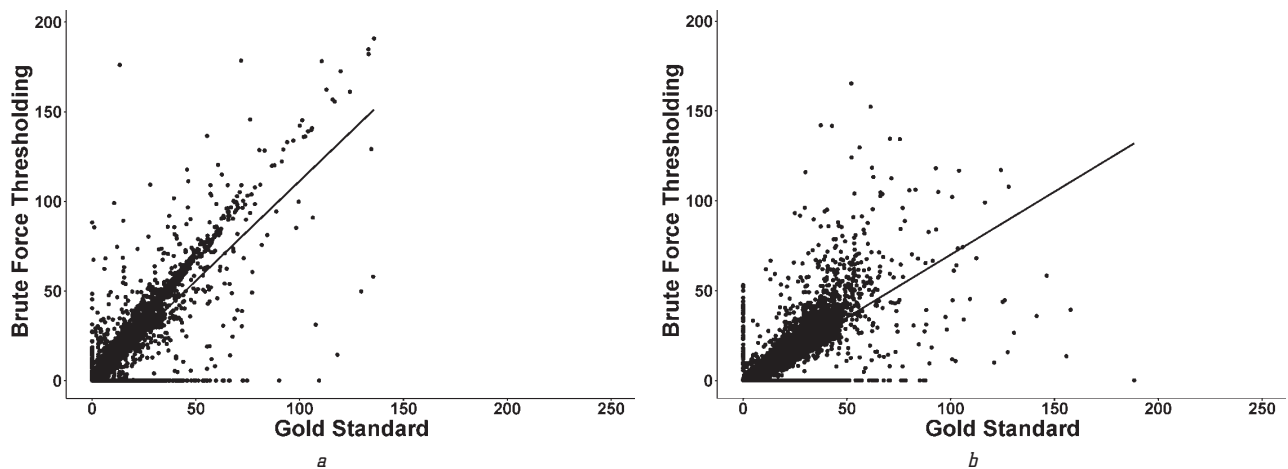

**Fig. 2.** Scatter plots of CBV perfusion maps that were produced from the threshold images and from the reference standard with linear regression: *a* – area under curve method; *b* – deconvolution method

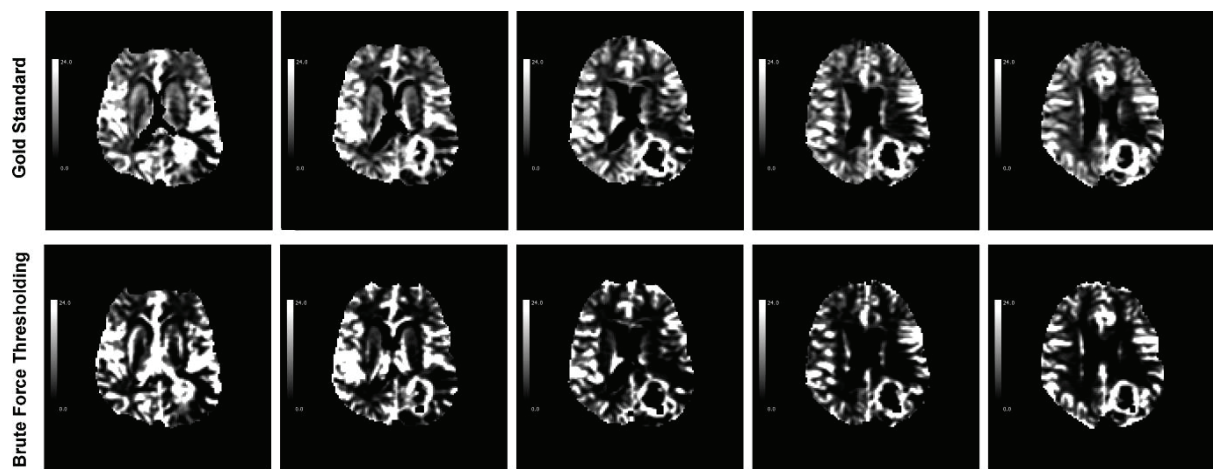

**Fig. 3.** CBV perfusion maps for a sample case produced with area under curve method (all maps are shown with the same window/level settings)

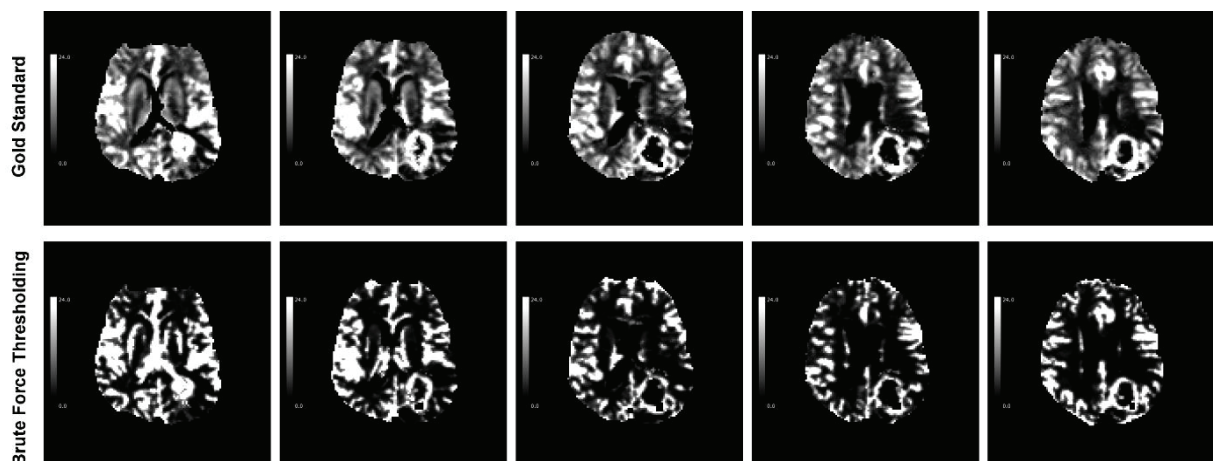

**Fig. 4.** CBV perfusion maps for a sample case produced with deconvolution method (all maps are shown with the same window/level settings)

This study has some limitations to be considered when interpreting the results. First, DSC perfusion MR datasets from only 12 patients were used. Further validation should be conducted by using a higher number of patients. Second, there are variations in the calculation algorithms of CBV values evaluation for both a mathematical-based semi-quantitative and physiological-based methods. The ability to produce more accurate results was the basis of the choice of weighted sum of the estimates of rectangle and trapezoid rule as algorithm for area under the curve method [14] and block-circulant singular value decomposition as algorithm for deconvolution method [4]. Because these other algorithms of CBV values evaluation were not directly validated in this study, further studies should be performed by comparing the results of different calculation algorithms. Third, since both accurate ROI analysis and histogram analysis of brain lesions use CBV map to determine the region of abnormality responsible for perfusion data analysis [15, 16], brain perfusion ROI detection may affect it determining and require further investigation.

## 4. Conclusions

The results obtained in this study indicate that thresholding is an ineffective way to detect brain perfusion ROI on DSC perfusion MR images of a human head. Control of pixels expressed the same intensity level is the main limitation of the thresholding technique that adversely impacts the results of both semi-quantitative and physiological-based evaluation of CBV values. Therefore, applying threshold-based tools in clinical practice can cause degradation of CBV maps quality and as a result can be harmful to patient management.

Standardization of perfusion ROI detection with further acceptance into validation protocols of new systems for perfusion data analysis can facilitate determining whether a particular system meets clinical requirements and can be transferred from research to industry.

*Alkhimova Svitlana, PhD, Department of Biomedical Cybernetics, National Technical University of Ukraine «Igor Sikorsky Kyiv Polytechnic Institute», Ukraine, e-mail: asnarta@gmail.com, ORCID: http://orcid.org/0000-0002-9749-7388*